\documentstyle[12pt,epsfig]{article}        
\newlength{\dinwidth}                       
\newlength{\dinmargin}                      
\def\lapprox{\lower .7ex\hbox{$\;\stackrel{\textstyle <}{\sim}\;$}}
\def\gapprox{\lower .7ex\hbox{$\;\stackrel{\textstyle >}{\sim}\;$}}

\def\GeV{{\rm GeV}}

\def\d{{\rm d}}

\def\lsim{\mathrel{\rlap{\lower4pt\hbox{\hskip1pt$\sim$}}
    \raise1pt\hbox{$<$}}}                
\def\gsim{\mathrel{\rlap{\lower4pt\hbox{\hskip1pt$\sim$}}
    \raise1pt\hbox{$>$}}}                
\begin{document}
\begin{flushright}
DTP/96/82
\end{flushright}
\vspace*{0.4cm}
\begin{center}  \begin{Large} \begin{bf}
Drell--Yan asymmetries at HERA-$\vec{{\rm \bf
N}}$\footnote{Contribution to the proceedings of the
``Future Physics at HERA''-workshop,
DESY, Hamburg, 1995/96.}\\
  \end{bf}  \end{Large}
  \vspace*{5mm}
  \begin{large}
T.~Gehrmann$^a$, W.J.~Stirling$^{a,b}$\\ 
  \end{large}
\end{center}
$^a$ Department of Physics, 
University of Durham,
South Road, Durham, UK \\
$^b$ Department of Mathematical Sciences, University of Durham,
South Road, Durham, UK \\
\begin{quotation}
\noindent
{\bf Abstract:}
The production of Drell--Yan pairs in the double polarized mode of
HERA-$\vec{{\rm N}}$ is studied at next--to--leading order in QCD. 
It is found  that a measurement of this observable
could yield vital information on the polarization of the light quark
sea in the nucleon.
\end{quotation}
\section{Introduction}

Structure functions measured in deep inelastic lepton--nucleon
scattering probe a particular combination of quark distributions in
the nucleon. The mere knowledge of these structure functions is
therefore insufficient for a distinction between valence and sea quarks
and for a further decomposition of 
the light quark sea into
different flavours. These are only possible  if
additional information from other experimental observables is taken
into account. 

Fits of unpolarized parton distributions (see for example~\cite{mrsg}) obtain
this information from two sources. The weak structure functions measured in
neutrino--nucleon scattering probe different combinations of parton
distributions than their electromagnetic counterparts. The inclusion of
these structure functions in a global fit can therefore constrain the
flavour structure of the unpolarized sea. A direct probe of the
antiquark distributions in the nucleon is given by the production of
lepton pairs in hadron--hadron collisions~\cite{dy}, the Drell--Yan
process.
It is in fact the inclusion of data from {\it both} processes in the
global fits that
leads to a precise determination of the distribution of antiquarks and
its flavour decomposition. 

Recent fits of polarized parton distributions~\cite{gs95,GRSV} 
have to rely entirely on the available data on the polarized structure
function $g_1^{p,d,n}(x,Q^2)$. The distinction of valence and sea
quark contributions to this structure function is possible to a
certain extent if additional information from sum rules is taken into
account. The flavour structure of the polarized sea is, however,
completely unknown at present. It seems rather doubtful that more
precise measurements of this structure function will be able to
provide more information on these two issues.

Polarized neutrino--nucleon scattering experiments will not be
feasible in the foreseeable future, although a measurement of polarized weak
structure functions may be possible from charged current
interactions at HERA~\cite{schaeferproc} if polarization in the
collider mode can be achieved. 

An experimental study of the polarized Drell--Yan cross section would
be possible with the HERA-$\vec{{\rm N}}$ experiment, operated with a
polarized proton beam on   a polarized nucleon target. We will examine
the prospects of such a measurement in this article. 

\section{The polarized Drell--Yan process} 

The production of lepton pairs in hadronic collisions can be
understood as annihilation of a quark--antiquark pair to a virtual
photon, which decays into a lepton pair of invariant mass $M^2$. 
The polarized and unpolarized cross sections for this process
are conventionally
defined to be~\cite{ratcliffe}
\begin{displaymath}
\d \Delta \sigma \equiv \frac{1}{2}
\,\left(\d\sigma^{++}-\d\sigma^{+-} \right), \qquad    
\d \sigma \equiv \frac{1}{2}
\,\left(\d\sigma^{++}+\d\sigma^{+-} \right),
\end{displaymath}  
where $(++)$ and $(+-)$ denote the configurations of aligned and
antialigned hadron spins. 

In the QCD--corrected parton model, these hadronic cross sections 
can be expressed as a convolution of parton--level coefficient functions with
the appropriate parton distributions:
\begin{eqnarray}
\lefteqn{\frac{\d [\Delta]  \sigma}{\d M^2}  =  \frac{4 \pi \alpha}{9 s M^2}
\int_{0}^{1} \d x_1 \d x_2 \d z \; \delta (x_1x_2z-\tau) 
\sum_q e_q^2 } \nonumber \\
& & \Bigg\{ 
\left\{\, [\Delta] q_1(x_1,M^2)\, [\Delta]
\bar{q}_2(x_2,M^2) + (1 \leftrightarrow 2) \, \right\} \left( [-]
\delta (1-z)   + \frac{\alpha_s(M^2)}{2 \pi}  
[\Delta] c_{q}^{DY} (z) \right) \nonumber \\
&& + \left\{\left( \, [\Delta] 
q_1(x_1,M^2)+[\Delta] \bar{q}_1(x_1,M^2)\right)\, [\Delta]
G_2(x_2,M^2) + (1
\leftrightarrow 2) \, \right\}   \frac{\alpha_s(M^2)}{2 \pi}
[\Delta] c_{G}^{DY} 
(z)\Bigg\}, \label{signloDY}
\end{eqnarray}
with the scaling variable $\tau=M^2/s$. The parton distributions in
the (not necessarily identical) hadrons are denoted by $f_{1,2}(x_i,M^2)$.

The next--to--leading order corrections to the unpolarized coefficient
functions have
been calculated in~\cite{dynlo}, and the polarized corrections are
given in~\cite{ratcliffe,kamal}. It turns out  that inclusion of these
corrections is crucial at fixed--target energies, as they 
contribute about 30\% of the total cross section. 
A fully consistent study of the Drell--Yan process at next--to--leading
order was until now only possible in the unpolarized case, as the polarized
parton distributions could only be determined to leading accuracy. 
With the recently calculated polarized two--loop splitting
functions~\cite{nlopol}, the polarized distributions can now be determined to
next--to--leading order from fits to structure function
data~\cite{gs95,GRSV}. 

\begin{figure}[t]
\begin{center}
~ \epsfig{file=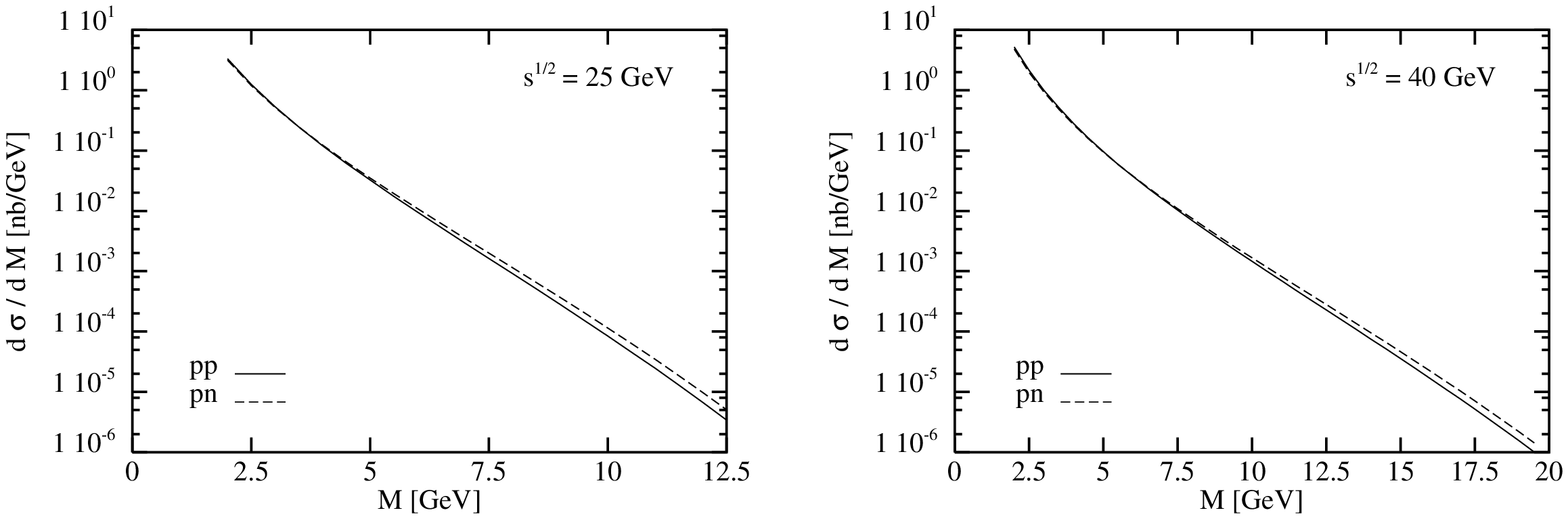,angle=-90,width=15cm}
\caption{{\it
Unpolarized Drell--Yan cross sections in proton--proton and
proton--neutron collisions.
  }}
\label{fig:DYu}
\vspace{0.5cm}
~ \epsfig{file=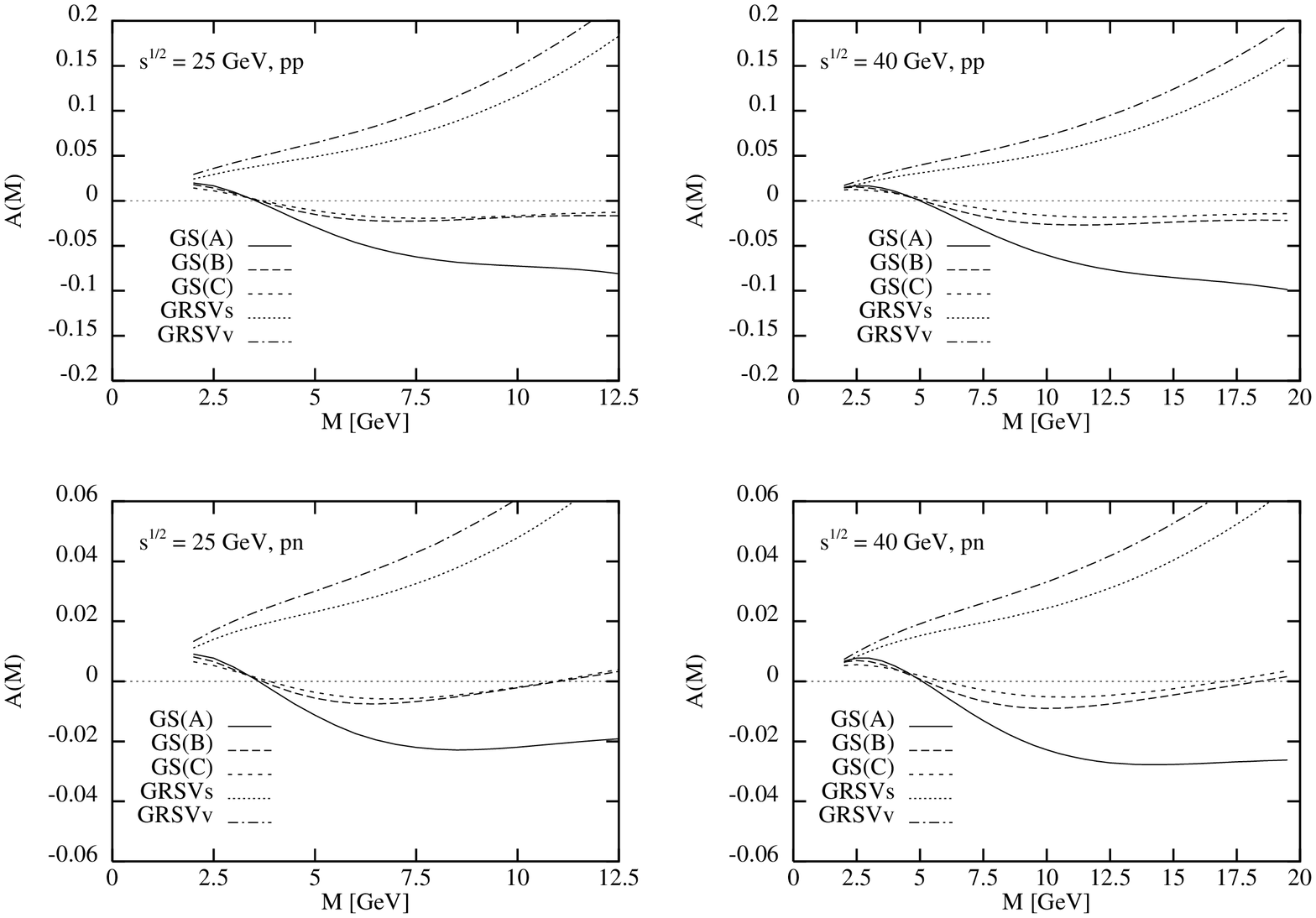,angle=-90,width=15cm}
\caption{{\it
Expected asymmetries in the polarized Drell--Yan process.
  }}
\label{fig:DYp}
\end{center}
\end{figure}

Using these distributions in combination with the unpolarized
distributions (set A$'$) from~\cite{mrsg}, we have calculated the total
Drell--Yan cross section $\d \sigma / \d M$
and the expected asymmetry 
\begin{displaymath}
A(M) \equiv \frac{\d \Delta \sigma / \d M}{\d \sigma / \d M}
\end{displaymath}
for proton and
(idealized) neutron targets at  centre--of--mass energies $\sqrt{s} =
40\;\GeV$ (HERA--$\vec{{\rm N}}$) and  $\sqrt{s} = 25\;\GeV$. The
latter could be achieved by operating HERA-$\vec{{\rm N}}$ with a
proton beam energy of about 330~GeV. 
Figure~\ref{fig:DYu} shows the
unpolarized Drell--Yan production cross section as a function of the
invariant mass of the lepton pair. It should be noted that invariant 
masses $M\le 4\;\GeV$ and $9\;\GeV \le M \le 11\;\GeV$ must be
excluded from the experimental measurement, as lepton pair production
in these mass regions is dominated by the decay of quarkonium resonances. 
An experiment with $\sqrt{s} = 25\;\GeV$ will clearly be restricted to
the invariant mass range $4\;\GeV < M < 9\;\GeV$; depending on the
available 
luminosity, a measurement for $M>11\;\GeV$ could be possible at $\sqrt{s} =
40\;\GeV$.

The Drell--Yan cross section at HERA--$\vec{{\rm N}}$ ($\sqrt{s} =
40\;\GeV$) is about two orders of magnitude larger than at RHIC--SPIN
($\sqrt{s} = 200\;\GeV$) when evaluated at fixed $\tau$. 

Figure~\ref{fig:DYp} shows the asymmetries obtained with the polarized
NLO parton distributions of~\cite{gs95} (GS(A,B,C)) and~\cite{GRSV}
(GRSVs,v). The spread in these predictions reflects the present lack of
knowledge on the behaviour of polarized sea quark  distributions in the
region $x>0.1$. Not even the sign of the asymmetry at large $M$ is predicted.
A sizable asymmetry of more than $\pm$10\% can be expected 
in proton--proton collisions; the asymmetry in proton--neutron
collisions is considerably smaller. 

We have checked the perturbative stability of these results
by varying the mass factorization scale; we find 
that the absolute value of the asymmetry is decreased (increased) by a maximum of
1.5\% if we take $\mu_F
= 2M$ $(\mu_F=M/2)$. This variation is significantly  smaller than the 
difference between the different parton distribution functions.

\section{Conclusions and Outlook}

A measurement of the polarized Drell--Yan cross section in the double
polarized mode of 
HERA-$\vec{{\rm N}}$ appears feasible, provided an
integrated luminosity of 100~pb$^{-1}$ or more can be achieved. 
Such a measurement would provide important information on the
polarization of the light quark sea at large $x$, a region which
cannot be probed with measurements of polarized weak structure
functions. Such a measurement would be unique to HERA-$\vec{{\rm N}}$,
as the polarized Drell--Yan process cannot be studied at the RHIC. 
Furthermore, HERA-$\vec{{\rm N}}$ could measure Drell--Yan asymmetries
off different targets, which could in principle be used to infer the flavour
structure of the polarized sea. Such a measurement would however
require much higher luminosity due to the small asymmetries on the
(idealized) neutron target.

In this study we have only examined the invariant mass distribution of the
Drell--Yan pairs, which is already able to discriminate different
parametrizations for the polarized sea quark distributions. Even more
information can be gained from more differential distributions (e.g.~in
the lepton--pair rapidity), which could be obtained with
higher luminosity.

\end{document}